\documentclass[useAMS,usenatbib]{mnras} 
\usepackage{amsmath}
\usepackage{graphicx}
\usepackage[T1]{fontenc}
\usepackage{epstopdf}
\graphicspath{{../Figures/}}

\title[]{The transformation and quenching of simulated gas-rich dwarf satellites within a group environment}
\author[C. Yozin and K. Bekki]{C. Yozin\thanks{{\bf E-mail} 21101348@student.uwa.edu.au; kenji.bekki@uwa.edu.au} and K. Bekki\footnotemark[1]
\\\\
ICRAR, M468, The University of Western Australia, 
35 Stirling Highway, Crawley
Western Australia, 6009, Australia}

\begin{document}
\date{Accepted 2013 January. Received 2013 January; in original form 2013 January}
\pagerange{\pageref{firstpage}--\pageref{lastpage}} \pubyear{2013}
\maketitle
\label{firstpage}

\newcommand{\asym}[1]{A$_{\rm #1}$}
\newcommand{\bt}{$B/T$}
\newcommand{\cia}{c$/$a}
\newcommand{\clr}{$^{\rm 1-0}$(g-r)}
\newcommand{\cmc}{cm$^{\rm -3}$}
\newcommand{\cnfw}{c$_{\rm NFW}$}
\newcommand{\cpet}{r$_{\rm 90}$/r$_{\rm 50}$}
\newcommand{\defhi}{def$_{\rm H {\sc I}}$}
\newcommand{\esf}{e$_{\rm SF}$}
\newcommand{\ex}[1]{10$^{\rm #1}$}
\newcommand{\fgas}{f$_{\rm gas}$}
\newcommand{\fkin}{f$_{\rm kin}$}
\newcommand{\ha}{H${\alpha}$}
\newcommand{\hi}{H~{\sc I}}
\newcommand{\htw}{H$_{\rm 2}$}
\newcommand{\kms}{kms$^{\rm -1}$}
\newcommand{\md}{M$_{\rm d}$}
\newcommand{\msol}{M$_{\odot}$}
\newcommand{\mdpc}{M$_{\odot}$pc$^{-2}$}
\newcommand{\mdyr}{M$_{\odot}$yr$^{-1}$}
\newcommand{\mex}[2]{#1$\times$10$^{\rm #2}$}
\newcommand{\mst}{M$_*$}
\newcommand{\mvir}{M$_{\rm vir}$}
\newcommand{\rars}{r$_{\rm H\alpha}$/r$_{\rm s}$}
\newcommand{\rgas}{r$_{\rm gas}$}
\newcommand{\rha}{r$_H{\alpha}$}
\newcommand{\rhoth}{$\rho_{\rm th}$}
\newcommand{\rp}{r$_{\rm p}$}
\newcommand{\rstr}{r$_{\rm str}$}
\newcommand{\rvir}{r$_{\rm vir}$}
\newcommand{\sd}{M$_{\odot}$pc$^{-2}$}
\newcommand{\sfr}{M$_{\odot}$yr$_{\rm -1}$}
\newcommand{\Sha}{$\Sigma_{\rm H_{\alpha}}$}
\newcommand{\si}{$\sim$}
\newcommand{\tp}{T$_{\rm p}$}
\newcommand{\tsn}{$\tau_{\rm SN}$}
\newcommand{\vcirc}{$V_{\rm circ}$}
\newcommand{\vsig}{$V/\sigma^*$}
\newcommand{\z}[1]{$z=#1$}


\newcommand{\figmap}
{
	\begin{figure}
	\includegraphics[width=1.\columnwidth]{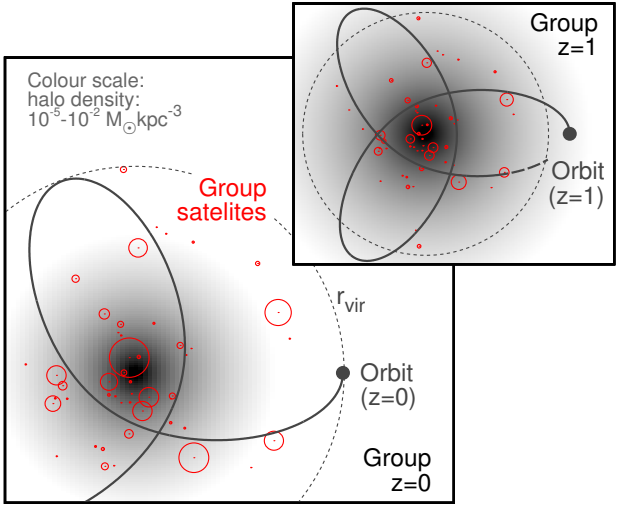} 
	\caption{Schematic of group models at \z{1} and \z{0}{} (left and right panels respectively). Black lines represent the satellite orbit in each case, with black circles denoting initial locations; red circles respresent the initial locations of other group satellites (size linearly correlated with satellite mass). In each case, the group halo density is conveyed in a logarithmic scale with the grey shaded region (darker$=$higher density).} 
	\end{figure}
}

\newcommand{\figlf}
{
	\begin{figure}
	\includegraphics[width=1.\columnwidth]{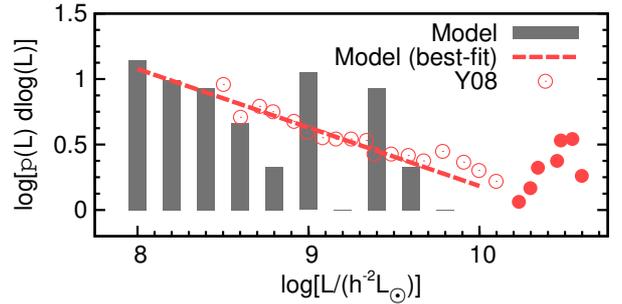} 
	\caption{Luminosity function for group satellites in the Tidal/\z{1}{}/Interactions simulations, shown in terms of binned luminosity (grey bars) and line-of-best-fit (red dashed line), compared with the equivalent data from SDSS DR4 \citep[][; Y08]{yang08} for groups in the mass range 12.9$<$log$_{\rm 10}$[M]$<$13.2, where open and closed circles denote satellite and central galaxies respectively } 
	\end{figure}
}

\newcommand{\figorbits}
{
	\begin{figure}
	\includegraphics[width=1.\columnwidth]{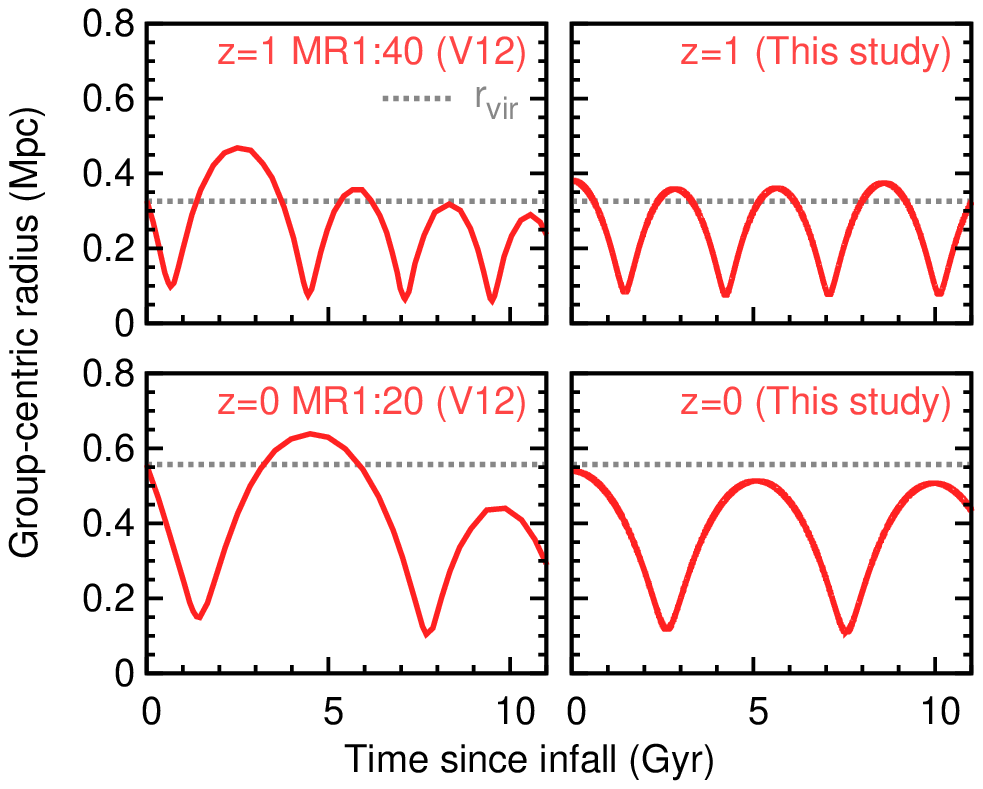} 
	\caption{A comparison of the orbits for a \ex{9}{} \msol{} satellite in a \ex{13}{} group (red solid line), where the grey dashed line represents the host virial radius. Left panels show the simulated orbit for this scenario from \citet{vill12} in which the satellite is accreted with the cosmologically most common trajectory and is subject to dynamical friction by a live host halo; right panels shows the equivalent orbit (i.e. with similar mean pericentre radii and orbital period) used in this work. Top and bottom panels show the \z{1}{} and \z{0}{} scenarios respectively.} 
	\end{figure}	
}

\newcommand{\figsflaw}
{
	\begin{figure}
	\includegraphics[width=1.\columnwidth]{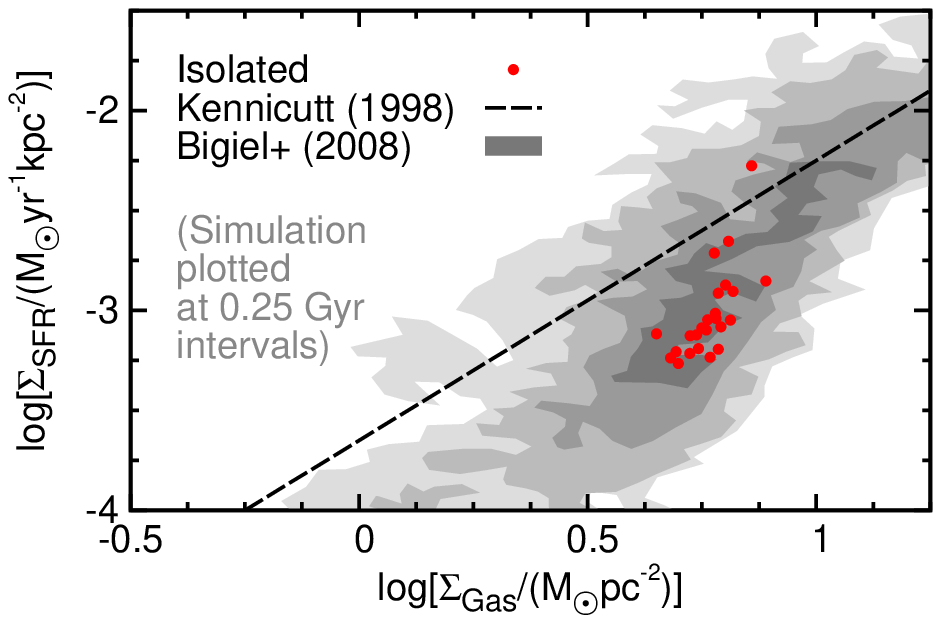} 
	\caption{Global (averaged) star formation law (\hi{}$+$\htw{} surface mass density vs. SFR surface density) for simulation Isolated (red circles), sampled at 0.25 Gyr intervals), compared with the Schmidt-Kennicutt law (power law index 1.4; black dashed line) for nearby normals and starburst spirals \citep{kenn98}, and 7 nearby spirals at 750 pc resolution (grey contours) reproduced from \citet{bigi08}.}
	\end{figure}
}

\newcommand{\figrotation}
{
	\begin{figure}
	\includegraphics[width=1.\columnwidth]{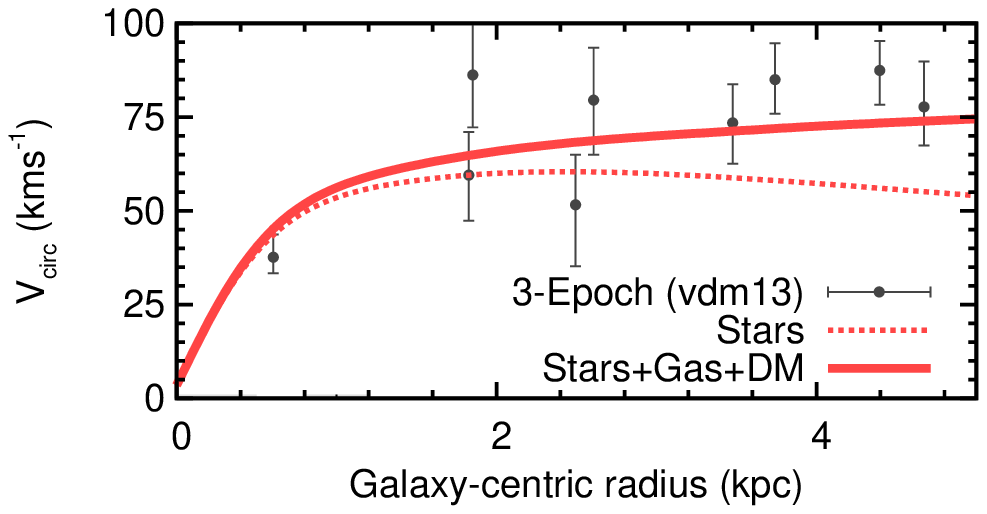} 
	\caption{Rotation curve for the disc model (red solid line), as inferred from $\sqrt{GM(<r)/r}$, with separate curve for the stellar (red dashed line) component. The model is compared to recent 3rd epoch HST data for the analogous LMC \citep[black error bars;][]{vdm13}.} 
	\end{figure}
}

\newcommand{\figsnapshots}
{
	\begin{figure*}
	\includegraphics[width=2.\columnwidth]{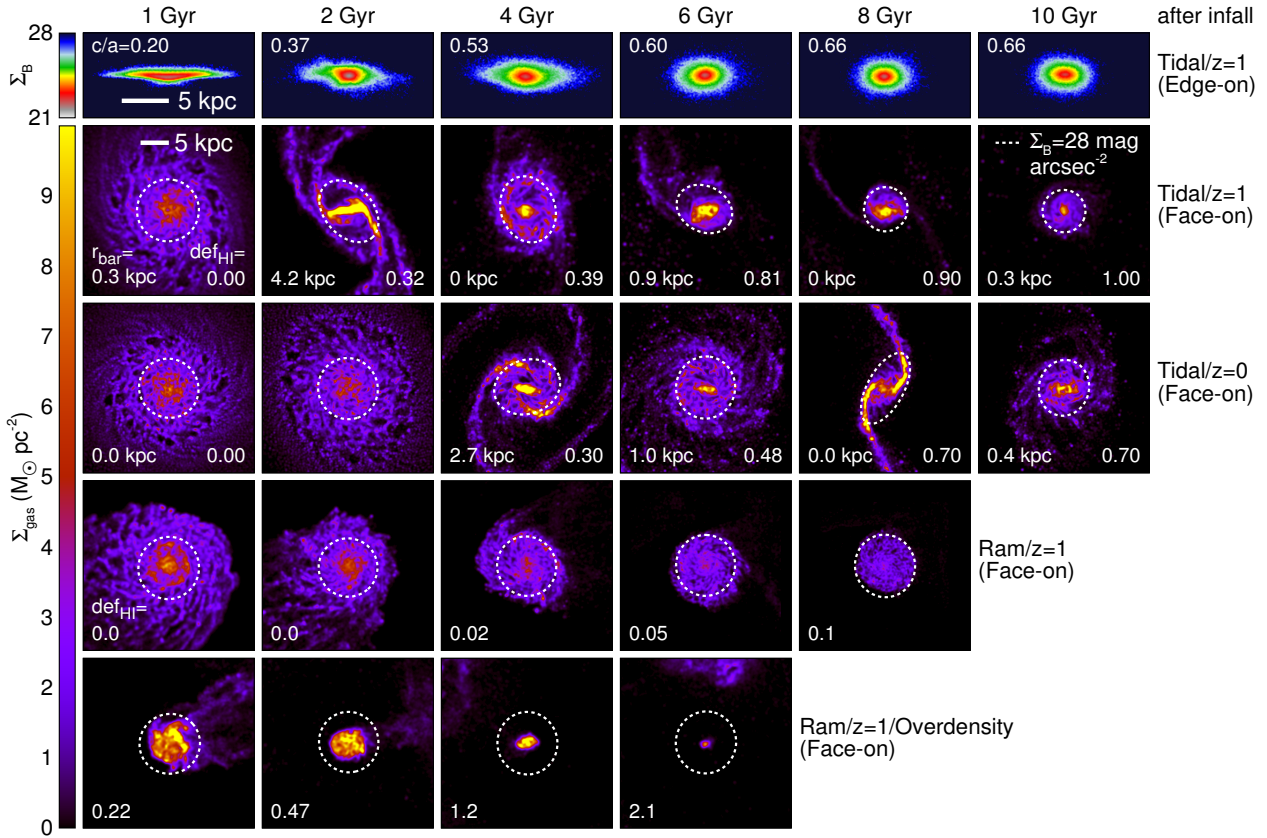} 
	\caption{Snapshots from simulations Tidal/\z{1}{}, Tidal/\z{0}{}, Ram/\z{1} and Ram/\z{1}{}/Overdense. The top row illustrates the $B$-band surface brightness of Tidal/\z{1}{}, viewed edge-on, to convey the thickening of the disc (with time since first infall labelled above, and the axial ratio c/a also provided). For all rows below, the underlying colour scale corresponds to the surface gas mass density, overlaid with an ellipse (white dotted line) fit to the 28 mag arcsec$^{\rm -2}$ isophote of the $B$-band surface brightness. Tidal simulations snapshots are provided with the instantaneous value of r$_{\rm bar}$, \asym{II} and \defhi{}; the bottom two rows show Ram simulations with the instantaneous \defhi{}.}  
	\end{figure*}
}

\newcommand{\figdiagref}
{
	\begin{figure*}
	\includegraphics[width=2.\columnwidth]{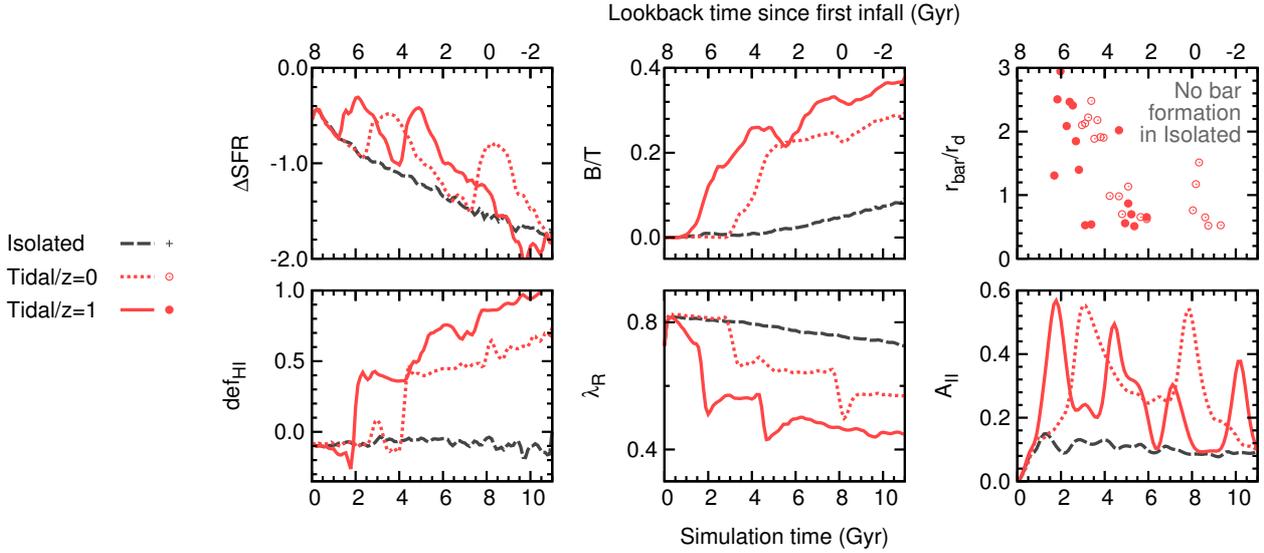} 
	\caption{Diagnostic of the evolution of simulations Tidal/\z{1} (solid red line), Tidal/\z{0} (dotted red line) and Isolated (dashed black line). (Clockwise from top left panel) Star formation rate (logarithmic) expressed as an offset from the expected SF for its stellar mass \citep[from][]{kauf03b}, where a passive or quenched classification is taken to lie at an $\Delta$SFR of -1 or -2 respectively; Bulge-to-total luminosity ratio (B/T); stellar bar radius r$_{\rm bar}$, normalised by the stellar scalelength r$_{\rm d}$; amplitude of the second Fourier mode \asym{II} averaged in radial bins up to the tidal radius, where the signature of spirals lie in the range 0.15 (weak arms) to 0.6 (strong); specific angular momentum of all stars within the effective disc radius $\lambda_{\rm R}$; \hi{}-deficiency of the galaxy, describing the \hi-mass within its optical diameter with respect to that expected for its morphological type as established from B/T. 
	}
	\end{figure*}
}

\newcommand{\figtoomre}
{
	\begin{figure}
	\includegraphics[width=1.\columnwidth]{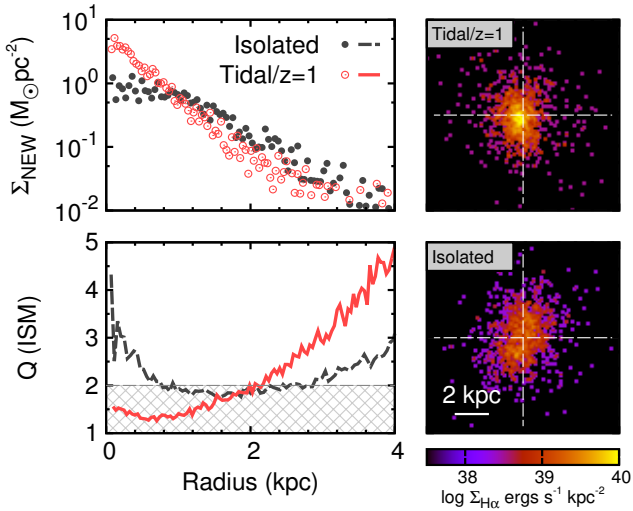} 
	\caption{(Left panels) Surface mass density of new stars (top) and Toomre stability parameter for the gas disc ($Q$) as a function of disc cylindrical radius, at 8 Gyr after infall, for Tidal\z{1}{} (red line/empty circles) and Isolated (black line/filled circle) simulations; (Right panels) H$\alpha$ emission density, in logarithmic scale, for the same simulations/time.
	}
	\end{figure}
}

\newcommand{\figbound}
{
	\begin{figure}
	\includegraphics[width=1.\columnwidth]{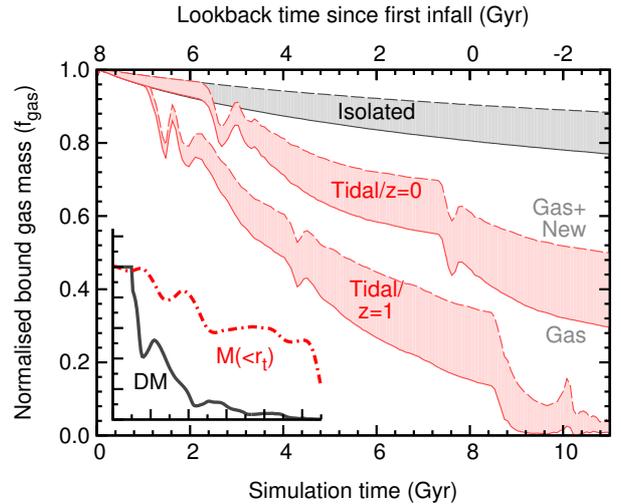} 
	\caption{(Main panel) For simulations Tidal/\z{1}{}, Tidal/\z{0}{} and Isolated, each band conveys the gas mass bound to the galaxy (lower limit) and the gas mass, plus stellar mass formed from gas, bound to the galaxy (upper limit), all normalised by the initial gas mass, as a function of simulation time; (Inset panel) the normalised DM mass bound in Tidal/\z{1}{} (black solid line), and the normalised total mass within the tidal radius (red dashed line).
	}
	\end{figure}
}

\newcommand{\figprofile}
{
	\begin{figure}
	\includegraphics[width=1.\columnwidth]{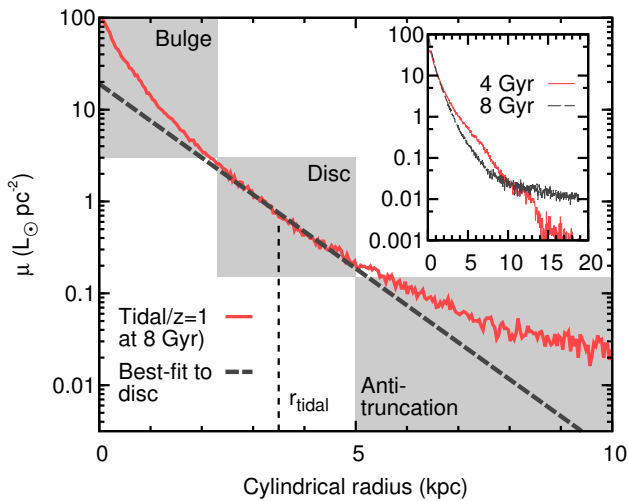} 
	\caption{
	(Main panel) Surface brightness profile, as a function of disc cylindrical radius, at 8 Gyr after infall for simulation Tidal/\z{1}{}. The best-fit line to the disc component is shown (black dashed line); the \bt{} ratio is computed from the excess light with respect to this line, lying within the tidal radius (r$_{\rm tidal}$); (Inset panel) a comparison of $\mu$-profiles for Tidal/\z{1}{} at 4 and 8 Gyr after infall.
	}
	\end{figure}
}

\newcommand{\figangmom}
{
	\begin{figure}
	\includegraphics[width=1.\columnwidth]{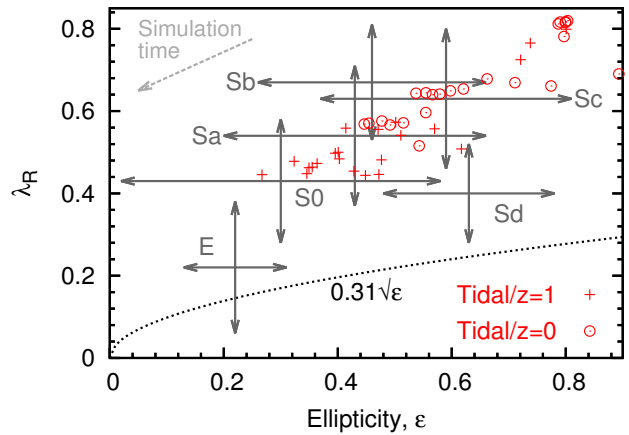} 
	\caption{
	Evolution of specific angular momentum (within r$_{\rm e}$) for simulations Tidal/z=1 and Tidal/z=0 (red plus and circle symbols respectively) as a function of stellar ellipticity. Points are plotted at 0.5 Gyr increments; as an idealised bulgeless disc, each simulation commences in the vicinity of ($\varepsilon=0.8$, $\lambda_{\rm R}=0.8$). Black bars illustrate the 1$\sigma$ distribution in each dimension for 300 CALIFA galaxies classified by morphological type \citep[reproduced from][]{falc14}. The grey arrow conveys the general evolution of our Tidal simulations; the dotted line is an effective demarcation between fast/slow rotating galaxies defined by $\lambda_{\rm R}=\sqrt{\varepsilon}$ \citep{emse11}.
	}
	\end{figure}
}

\newcommand{\figdiagparams}
{
	\begin{figure*}
	\includegraphics[width=2.\columnwidth]{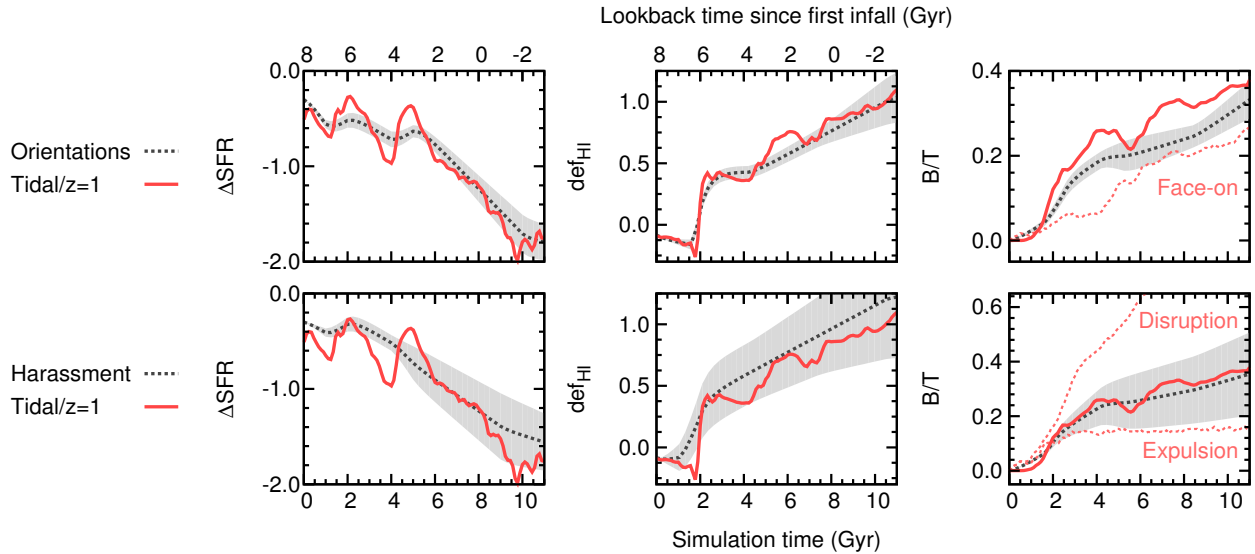} 
	\caption{Diagnostic of $\Delta$SFR, \defhi{}, and \bt{} (as in Fig. 7), shown for simulation Tidal/\z{1}{} (red solid line), and the mean/1$\sigma$ variation (black dotted line/grey shaded region) in the respective evolution of two sets of simulations run in accord with Monte-Carlo sampling, where the top row shows the results for 16 variants of Tidal/\z{1} where the initial orientation of the satellite (with respect to its orbital plane) is assigned randomly (Orientations), and the bottom row shows the results for 16 variants of Tidal/\z{1}{} where other group galaxies are incorporated in the model as point masses and the primary satellite is assigned a random initial location on the locus defined by the group halo \rvir{} (Harassment). In the B/T panels, the evolution of selected individial simulations from each set are shown, including a satellite accreted face-on in inclination, and two simulations in which harassment strongly alters the primary satellite's orbit and thus its transformation process (red dotted lines).} 
	\end{figure*}
}

\newcommand{\figram}
{
	\begin{figure}
	\includegraphics[width=1.\columnwidth]{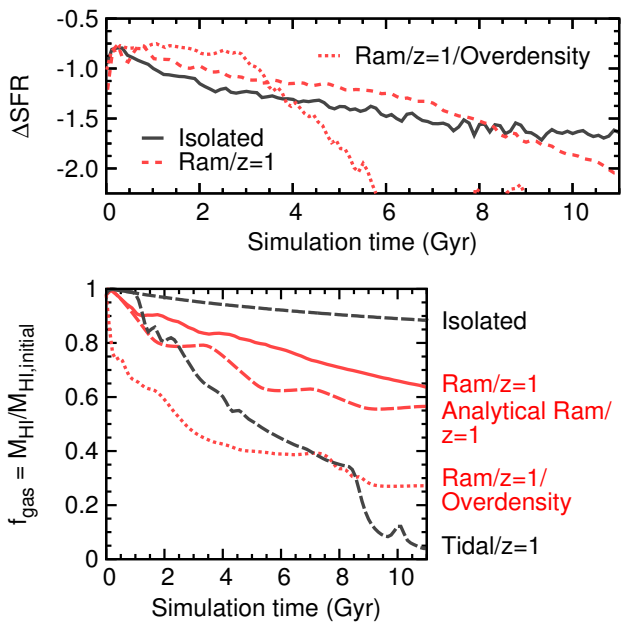} 
	\caption{(Top panel) $\Delta$SFR for simulations Isolated, Ram/\z{1}{} and Ram/\z{1}/Overdensity; (bottom) \fgas{} ($=$M$_{\rm HI}$/M$_{\rm HI,initial}$) for simulations Isolated, Ram/\z{1}, Ram/\z{1}/Overdensity and Tidal/\z{1}{}. Also shown is an analytical estimate of \fgas{} (black dotted line), using the approximation of r$_{\rm str}$ for the \z{1}{} orbit and assuming an exponential cold gas disc.}
	\end{figure}
}


\newcommand{\tablegroup}
{
	\begin{table} 
	\centering
	\caption{Summary of group model and orbital parameters}
	\begin{tabular}{@{}lcc@{}}
	\hline
	Parameter & Value at \z{1} & Value at \z{0}{} \\
	\hline
	Virial mass (\msol) & \ex{13}{} & \ex{13}{} \\
	Virial radius (kpc) & 329 & 556 \\
	\cnfw{} & 4.87 & 9.74 \\
	Apocentre (\rvir) & 1.2 & 1.0 \\
	Pericentre (\rvir) & 0.24 & 0.2 \\
	Pericentre Velocity (kms$^{\rm -1}$) & 650-700 & 550-600 \\
	\end{tabular}
	\end{table} 
}

\newcommand{\tablemodel}
{
	\begin{table} 
	\centering
	\caption{Summary of model parameters}
	\begin{tabular}{@{}lr@{}}
	\hline
	Parameter & Value \\
	\hline
	DM particle Number & \mex{1}{6}  \\
	DM-to-stellar mass ratio (M$_{\rm h}$/\msol{}) & 100 \\
	NFW concentration (c$_{\rm NFW}$) & 7-12 \\
	Virial radius (r$_{\rm vir}$/kpc) & 120 \\
	\hline
	Stellar particle Number & \mex{3}{5}  \\
	Stellar Mass (M$_{\rm d}$/\msol{}) & \ex{9} \\
	Stellar scalelength ($r_{\rm d}$/kpc) & 1.1 \\
	Stellar scaleheight ($z_{\rm d}$/kpc) & 0.22 \\
	Toomre parameter (Q) & 2.0 \\
	Stellar metallicity ([Fe/H]) & -0.9 \\
	Metallicity gradient (d[Fe/H]/d$r$/dex/kpc) & -0.04 \\
	SNe expansion timescale (t$_{\rm SN}$/yr) & \ex{5.5}{} \\
	\hline
	Gas particle Number & \mex{2}{5} \\
	Gas mass fraction (f$_{\rm gas}=M_{\rm g}/M_{\rm d}$) & 1.0 \\
	Gas scalelength (r$_{\rm g}$/r$_{\rm d}$) & 2.6 \\
	Gas scaleheight (z$_{\rm g}$/z$_{\rm d}$) & 2.0 \\
	SF Threshold Density ($\rho_{th}$/cm$^{\rm -3}$) & 10$^{\rm 0}$ \\
	\end{tabular}
	\end{table} 
}

\newcommand{\tablesims}
{
	\begin{table}
	\centering
	\caption{Summary of simulations}
	\begin{tabular}{@{}lp{40mm}@{}}
	\hline
	Simulation & Description \\
	\hline
	Isolated & Isolated model (not orbiting within group potential) \\
	Tidal/\z{1}{} & Satellite under tidal influence from the group potential only; orbit and group properties at \z{1}{} \\
	Tidal/\z{0}{} & As above, but with orbit and group properties at \z{0}{} \\
	Ram/\z{1}{} & Satellite under Ram Pressure only; properties at \z{1}{} \\
	Ram/\z{1}{}/Overdense & As above, but with IGM density increased by factor 10 \\
	Tidal/\z{1}/Orientations & Tidal/\z{1}{}, but with 16 variations on the initial orientation w.r.t orbital plane \\
	Tidal/\z{1}/Harassment & Tidal\z{1}{}, but with other group satellites included, 16 variations on initial position \\
	\end{tabular}
	\end{table}
}

\begin{abstract}

The underlying mechanisms driving the quenching of dwarf-mass satellite galaxies remain poorly constrained, but recent studies suggest they are particularly inefficient for those satellites with stellar mass 10$^{\rm 9}$ M$_{\odot}$. We investigate the characteristic evolution of these systems with chemodynamical simulations and idealised models of their tidal/hydrodynamic interactions within the 10$^{\rm 13-13.5}$ \msol{} group-mass hosts in which they are preferentially quenched. Our fiducial simulations highlight the role played by secular star formation and stellar bars, and demonstrate a transition from a gas-rich to passive, \hi-deficient state (i.e. $\Delta$SFR$\le$-1, \defhi{}$\ge$0.5) within 6 Gyr of first infall. Furthermore, in the 8-10 Gyr in which these systems have typically been resident within group hosts, the bulge-to-total ratio of an initially bulgeless disc can increase to 0.3$<$B/T$<$0.4, its specific angular momentum $\lambda_{\rm R}$ reduce to $\sim$0.5, and strong bisymmetries formed. Ultimately, this scenario yields satellites resembling dwarf S0s, a result that holds for a variety of infall inclinations/harassments albeit with broad scatter. The key assumptions here lie in the rapid removal of the satellite's gaseous halo upon virial infall, and the satellite's local intra-group medium density being defined by the host's spherically-averaged profile. We demonstrate how quenching can be greatly enhanced if the satellite lies in an overdensity, consistent with recent cosmological-scale simulations but contrasting with observationally-inferred quenching mechanisms/timescales; an appraisal of these results with respect to the apparent preferential formation of dS0s/S0s in groups is also given.

\end{abstract}


\section{Introduction}


Besides their initial and often rapid formation, the evolution of most galaxies can be defined by a gradual suppression of star formation (quenching), usually coincident with a transformation from a disc to elliptical morphology, much of which is observed to occur within within cosmologically overdense regions like groups and clusters \citep{butc84, dres97, lewi02, kauf03a, balo04, wein06}. 

In the context of the $\Lambda$CDM framework, these observations partially reflect the early hierarchical assembly of massive galaxies via merging and their quenching via radio-mode AGN feedback \citep{balo04,bluc14}. This mass-quenching appears to dominate at high-$z$ and for masses above a characteristic \ex{10}{} \msol{} \citep{peng12, wetz13}. The (dwarf) galaxies that lie below this threshold are rarely found quenched in the field \citep{geha12}, but tracing their recent quenching/transformation to specific mechanisms as satellites remains difficult.

The main mechanisms are speculated to include tidal shock/heating by a host potential well \citep{maye01, kaza11}, harassment by successive tidal encounters \citep{moor98, wetz10, bekk11}, and hydrodynamical interactions with the hot gas halo of a host (Intra-cluster medium; ICM). The latter, often invoked to explain a suppression of star formation (SF), include in particular the ram pressure stripping of a satellites' cold gas disc \citep{gunn72, koop04, chun07} and the stripping of the satellites' hot gaseous corona that would otherwise fuel star formation \citep[strangulation;][]{lars80}.

The efficiency of these processes within a given environment is expected to vary with properties of the constituent baryons (i.e. satellite velocity dispersion, ICM temperature) and host halo. In practice, a bimodal (red and blue) galaxy population is observed in most overdense regions \citep{vdb08, cibi13, bluc14} since \z{0.2-0.5}{} \citep{mcge11, pope12}. This bimodality extends down to the host mass scale of \ex{13}{} \msol{}, in which the majority of all galaxies reside \citep{tuck00, eke04, rasm12}.


Recently, statistical analyses of large cosmological simulations and surveys (e.g. SDSS) have been adopted to establish i) the timescale of satellite quenching/transformation, with respect to ii) the time since accretion to an overdense region \citep[e.g.][]{delu12}. The finding that these timescales lie on the order of several Gyr is consistent with a slow quenching facilitated by strangulation \citep{mcca08, wein10}, as contrasted with the relatively rapid action of ram pressure stripping, which in some models produces too many passive satellites \citep{wang07}. 

Other studies have gone further to suggest that the quenching timescale is independent of host halo masses above $>$\ex{12-13}{} \msol{} \citep[e.g.][]{wetz13}, where any apparent variability corresponds instead to the hierarchical growth of such environments. In the event of this pre-processing, previously quenched satellites may be accreted as part of their host to a more massive host \citep{zabl98, mcge09, peng12}. This remains far from conclusive, however, with numerical studies differing in their prediction for the onset of these quenching mechanisms and their subsequent efficiency \citep{bekk09, mcca08, wetz13, bekk14, cen14, bahe15}. 

The uncertainty concerning environmental processing motivates a study in which we simulate in detail the fundamental tidal and hydrodynamical mechanisms acting on a dwarf satellite. Specifically, we consider the apparent inefficiency with which \ex{9}{} \msol{} satellites are environmentally-quenched \citep[][and references therein]{whee14}. By comparing the observed quenched fraction of these systems \citep{geha12} with their associated fraction residing as subhalos in a more massive host (as deduced with a mock catalog constructed from the Millenium II Simulation), \citeauthor{whee14} find that up to only 30 percent are quenched by their environment (most commonly a \ex{13-13.5}{} \msol{} host). Using simple analytical models, they hypothesize that the corresponding quenching mechanism must act only on long timescales (e.g. $\geq7.5$ Gyr).

This inefficiency contrasts with the distinct transition from blue irregulars to red early-types acknowledged amongst the smallest of the (\ex{6-9}{} \msol{}) dwarf galaxies when crossing the virial radius of the Local Group's largest galaxies \citep{mcco12}. While ram pressure interactions with the Galactic hot halo and cosmic reionisation have been demonstrated to reproduce their current paucity of detectable gas \citep{maye01}, the tidal stirring mechanism (the repeated tidal shock/heating of a satellites stellar disc) reveals the efficiency with which groups environments can transform very faint dwarf galaxies \citep{maye01, kaza11} and even Galactic-mass systems \citep{bekk11, vill12}. Group mechanisms also appear to be responsible for the dramatic number growth, since at least \z{0.5}{}, of the quenched and early-type systems known as Lenticular (S0) galaxies \citep{dres97, will09, just10}.

In this analysis, we adopt methods demonstrated in our previous studies \citep{bekk11, yoz14a, bekk14} and the results of previous cosmological studies that predict the infall epochs and group orbits for this stellar mass regime \cite[e.g.][]{delu12, vill12}, and devise an idealised model of this \ex{9}{} \msol{} satellite and a \ex{13}{} \msol{} host system to characterise its long-term evolution. 

Other studies have considered the tidal transformation of dwarves within groups/clusters \citep{smit10, kaza11, vill12, bial15}, and the star formation histories of Galactic-mass satellites in groups \citep{feld11}. Besides our novel choice of satellite and group masses, we distinguish this work with i) an estimation of the SF history and \hi-deficiency using our chemodynamical model of the interstellar medium (ISM); ii) an interpretation of our results in terms of the stochasticity/evolutionary scatter introduced by infall inclinations/satellite interactions/group mass assembly, and iii) complementary simulations of hydrodynamical (ram pressure) interactions between the group medium and satellite. 

Section 2 describes the group and satellite models, while in Section 3, we illustrate the results of a suite of the tidal/hydrodynamical simulations. A summary and interpretation of these results in the context of previous observational/numerical studies is given in Section 4.

\section{Method}

We model satellite galaxy infall into a \ex{13}{} \msol{} halo to assess the efficiency/mechanisms by which it is transformed and quenched. This involves placing an idealised dwarf galaxy (pristine bulgeless disc) at or beyond the virial radius (\rvir{}) of a static group potential, with a trajectory prescribed from previous studies on the most cosmologically-common infall orbit. This evolution is assessed from a lookback time of 8 to 10 Gyr, which represents the median epoch at which \ex{9}{} \msol{} satellites resident in group-mass hosts (\ex{13}{} \msol{}) at \z{0}{} were first accreted from isolation \citep{mcge09, delu12}. A schematic of this scenario is illustrated in Fig. 1.

This method permits us to disentangle the primary mechanisms presently believed to govern dwarf satellite evolution, namely group tides, harassment, ram pressure and secular evolution. Crucially, we utilise satellite models with a \ex{3-4}{}-fold improvement on baryonic resolution with respect to the cosmological-scale simulations whose results motivated this study. However, our method has disadvantages insofar as the group model neither follows the mass growth (by factor 2) expected from hierarchical growth over this timescale \citep{fakh10}, nor allows for the potential merging and/or resonant stripping of jointly accreted satellites \citep{gned03, dong09}. We must further assume the instantaneous stripping of the satellites' extended gas reservoir upon first crossing the group \rvir{}, such that this halo can be excluded from our modelling. This choice is motivated by the explicit hydrodynamic modelling of this process in galaxy-scale simulations \citep{kawa08} and other studies which suggest the relatively dffuse intragroup medium (IGM) can even remove a satellites' cold gas disc \citep{bahe15}. 

In the following sections, the group/galaxy models and simulations are described in further detail; we discuss our modelling choices in Section 4.2.

\subsection{Group Model}

\figmap
\tablegroup

Our group model is built on the assumption of a virialized group environment in which most substructure is accreted into a spherically symmetric background halo. For simplicity, both the dark halo and IGM are modelled with a NFW density profile \citep{nava96}, which statistical analyses of clusters suggest provides an acceptable mass distribution \citep{vdm00}: 
\begin{equation}
\rho(r)\propto\frac{1}{r/r_{\rm group}(1+r/r_{\rm group})^{\rm 2}},
\end{equation}
where $r$ is the group-centric radius and $r_{\rm group}=r_{\rm vir}$/c$_{\rm NFW}$ where c$_{\rm NFW}$ is the concentration factor. 

To overcome the idealistic nature of our static group model, we adopt two representations of a group at distinct epochs of \z{1}{} and \z{0}{}, whose combined results encompass the characteristic evolution of a dwarf satellite accreted since \z{1}{}. We fix the group mass to \ex{13}{} \msol{} at both epochs, and extract c$_{\rm NFW}$/\rvir{} from the similar models of \citet{vill12}; the adopted properties are summarised in Table 1.

\figlf

For the subset of simulations in which we explicitly add other group galaxies, we adopt the method introduced in \citet{bekk11} in which they are modelled as collisionless point masses, orbiting freely within the group, with an interaction softening length fixed to half each satellites' \rvir{}. The total mass of galaxies M$_{\rm gr}$ is established according to the total luminosity L$_{\rm gr}$, with the relation of \citet{mari02}:
\begin{equation}
	\frac{M_{gr}}{L_{gr}} = 350\left({\frac{M_{gr}}{5\times10^{14}M_{\odot}}}\right)^{0.335}.
\end{equation}
Monte-Carlo sampling is adopted to fit satellites, limited to a luminosity range 0.01 to 2.5L$^*$ and with an assumed mass-to-light ratio of 40, to the Schechter function with a faint-end slope -1.07 consistent with groups as established by \citet{yang08}. Thus, our \ex{13}{} \md{} group hosts 52 satellites, with an initial spatial distribution prescribed with a NFW profile and a concentration of 3.0 to match the galaxy distributions revealed in K-band studies \citep{lin04}. Fig. 2 demonstrates good agreement between the corresponding linear fit to our LF with that obtained for groups in the mass range 12.9$<$log$_{\rm 10}$[M]$<$13.2 from the SDSS DR4 \citep{yang08}. The corresponding velocity dispersion of these satellites, a proxy for the influence yielded by satellite-satellite tidal encounters, is consistent with observations at between 125 and 150 \kms{}.

\subsection{Orbits}

\figorbits

We derive the orbits from those exhibited by a $\sim$\ex{9}{} satellite in the pure tidal/collisionless simulations of \citet{vill12}, who incorporated a live group halo to model its capacity to retard the satellite via dynamical friction. A satisfactory live model of this halo requires significant resolution to avoid discreteness effects (a resource we choose instead to dedicate to the satellite model); given that analytical methods are not sufficiently accurate \citep{jian08} and that this drag is nominally weak for dwarf satellites \citep{smit10}, we exclude this friction in our model. 

Fig. 3 conveys how we estabilish from \citeauthor{vill12} an elliptical orbit at epochs \z{1}{} and \z{0}{} from the mean pericentres and apocentres across an 8-10 Gyr interval since first crossing the group \rvir{}. In each case, the apocentre-to-pericentre ratio is consistent with the median among large-scale simulations \citep[i.e. 4-5][]{ghig98}. Qualitatively, it can be seen that each orbit differs significantly in its pericentre radii and orbital period; given that each is smaller in the \z{1}{} case, we expect its satellite evolution to be most dramatic. In each case, we deem the orbit sufficiently far from the group centre not to warrant the explicit modelling of a massive central galaxy often found lying at the X-ray temperature peak among groups \citep{zabl98}, and responsible for potentially efficient tidal shocks \citep{maye01}.

\subsection{Disc Model}

\figrotation

\tablemodel

Table 2 summarises the initial conditions of our satellite, a gas-rich bulgeless disc, which is comprised of a dark halo and stellar/gaseous disks. All galaxy models are comprised of \mex{1.5}{6} particles, divided between halo, stellar and gas components with \ex{6}, \mex{3}{5} and \mex{2}{5} particles respectively.

For the stellar disc mass M$_{\rm d}=10^{\rm 9}$ \msol{}, we find a halo mass M$_{\rm h}\simeq100{\rm M}_{\rm d}$ from observed intermediate-$z$ dwarf galaxies and abundance matching within large-scale simulations \citep{mill13, muns13, behr13}, but note that the latter may produce too much substructure, in which case our M$_{\rm h}$ may be a slight overestimate. We assume a top hat collapse model in which the mean halo density enclosed within \rvir{} reflects the critical density at formation, and therefore use a scaling relation of the form ${\rm r}_{\rm vir}\propto{\rm M}_{\rm h}^{\rm 1/3}$ normalised by properties of the Galaxy \citep{klyp02}, from which we establish a satellite \rvir{} of 120 kpc. We utilise a NFW density profile for the halo; the single distinction between our satellite at \z{1}{} and \z{0}{} lies in its initial concentration \cnfw{}, which we constrain from mass-redshift-dependent relations of \citet{muno11}, and establish an evolution of \cnfw{} since \z{1}{} from 7 to 12.

An exponential disc morphology is adopted for the satellite, (truncated at 5 r$_{\rm d}$) with a density defined for a cylindrical radius $r$ and height $z$:
\begin{equation}
	\rho(r,z)\propto{\rm exp}\langle{-}\frac{r}{r_d}\rangle{\rm sech}^{\rm 2}\langle\frac{z}{0.2r_d}\rangle,
\end{equation}
where r$_{\rm d}$ is a stellar scalelength derived from size-mass scaling relations \citep{ichi12} which exhibit a power law (index $\alpha=0.15$). This differs from simple self-similar models of galaxy formation \citep[e.g.][]{moma98} which do not account for mass-dependent processes such as outflow. Disc kinematics are constrained on the basis that the typical late-type disc with this M$_{\rm d}$ will not host a secular bar instability \citep[i.e.][]{melv13}. Initial velocity dispersions, assigned according to epicyclic theory (with the vertical dispersion at a given radius is half that of the associated radial velocity dispersion), are thus chosen with a Toomre's parameter of $Q=2$ to inhibit expedient axisymmetric collapse \citep{comb81}. 

We enforce a gas-to-stellar mass ratio of unity from semi-empirical scaling laws for the gas mass (as a function of $z$) based on indirect measures of \hi{} and \htw{} content (e.g. the SF rate), which compares well with the high observed gas fractions of dwarves \citep[][and references therein]{popp14}. We assume this mass is also distributed with the aforementioned exponential profile, albeit with a gas scalelength r$_{\rm g}=2.6r_{\rm d}$ as established from the normalised galaxy sample of \citet{krav13}, and an assumed scaleheight z$_{\rm g}=2z_{\rm d}$.

Fig. 4 illustrates how the initial rotation curve of this model fits acceptably to that of the LMC \citep{vdm13}, an exemplary galaxy in this stellar mass regime.

\subsection{Numerical modelling}

We utilise an original parallelised chemodynamical code GRAPE-SPH \citep[GRAvity PipE-Smoothed particle hydrodynamics;][]{bekk09}. Gravitational dynamics are computed with a softening length for halo and stellar particles ($\eta_{\rm h}$$=$2.1 kpc and $\eta_{\rm s}$$=$200 pc respectively) which is set by the mean particle separation at the halo and stellar half-mass radii respectively. The ISM is modelled with smoothed particle hydrodynamics (SPH), with a minimum smoothing length of $\sim$30 pc, and can cool radiatively \citep[see][for more details]{bekk14}. This depends on an initial galaxy-wide mean metallicity of $[Fe/H]=-0.9$ prescribed from mass-metallicity relations \citep{kirb13}, and a fixed in-plane radial metallicity gradient of 0.04 dex kpc$^{\rm -1}$. 

Star formation (SF) occurs via the conversion of gas particles to new stellar particles; this process is limited to collapsing Jeans-unstable regions with volumetric density exceeding a threshold \rhoth{}, and relies on a probability parameter, P$_{\rm SF}$, introduced in \citet{bekk13}:
\begin{equation}
P_{SF}=1-exp(-\Delta t \rho^{\alpha_{sf}-1}),
\end{equation}
where $\Delta t$ is the timestep width for a given gas particle. If P$_{\rm SF}$ exceeds a randomly generated number $R$ with $0\le{\rm R}\le1$, the conversion proceeds, based on the assumption that the timescale for star formation is shorter for higher density gas particles. This condition softens the numerical limitation that the giant molecular clouds which host SF lie below the resolution of gas particles. 

\figsflaw

Stellar feedback is implemented with a sub-grid Supernova (SN) model \citep[see][for more details]{bekk14} in which each subgrid SN type II expels a canonical $10^{51}$ ergs, with 90 percent delegated to thermal UV emission which is conferred upon neighbouring gas particles over a prescribed adiabatic expansion timescale \tsn{}. The rate at which SNe occur is governed by an initial mass function (IMF), for which we adopt the standard Salpeter power law ($\alpha=-2.35$) normalised to within the range 0.1 M$_{\odot}$ and 100 M$_{\odot}$.

This ISM model is simple in terms of its IMF and the lack of an explicit modelling of \htw{}-formation/outflow. However, in Fig. 5 we illustrate how our choice of free parameters \rhoth{} and \tsn{} (1 \cmc{} and \ex{6}{} yr respectively) can sufficiently replicate the observed SF law of dwarf galaxies \citet{bigi08}, who adopt far UV and 24$mu$m emission to trace the inefficient (relative to the Kennicutt relation) SF in 7 nearby spirals. We also ameliorate the concern that weak feedback schemes permit the existence of unphysical dark matter cusps, with the finding of a flat density profile within 1-2 percent of the satellite \rvir{} \citep[e.g.][]{broo12}.

For the subset of simulations in which we exclusively model the interactions between the satellite ISM and group IGM (i.e. no group potential), we adopt a method introduced in \citet{bekk14} in which the self-gravitating galaxy model is placed within a cubic lattice of SPH particles (side length 25r$_{\rm d}$) that represents the IGM local to the satellite. The galaxy model is fixed, such that variations in the relative velocity and density of its surrounding IGM are applied by modifying at each time step the properties of the IGM SPH particles, according to the satellite's orbit. Therefore, each IGM particle is assigned a density of f$_{\rm ICM}$$\rho$($r$), where $r$ is the group-centric radius, $\rho$ is the same NFW profile adopted to describe the group halo potential, and f$_{\rm ICM}$ is the mass quotient of the ICM with respect to the group halo. A fixed gas temperature of T$_{\rm ICM}=5.6\times10^{\rm 6}$ K is obtained for a \ex{13}{} \msol{} group from \citet{mats00}.

\subsection{Simulations}

\tablesims

This study analyses the results of 37 self-consistent simulations, whose details are summarised in Table 3. As a baseline case, we analyse the tidal influence of the group spherical potential at \z{1}{} and \z{0}{} (labelled Tidal), with respect to the pure secular evolution of an identical galaxy (Isolated). 

Simulations labelled Ram refer to those in which a self-gravitating galaxy model interacts only hydrodynamically with the IGM. We conduct two simulations to address the variations in the local density experienced by infalling satellites. The first, Ram/\z{1}{}, adopts the \z{1}{} group properties and a f$_{\rm ICM}$ of 0.1 consistent with the mean group ICM mass inferred from X-ray observations \citep{lin12}. In this case, our adopted ICM density profile compares sufficiently well with those which fit the distribution of galaxies, as inferred for example from X-ray temperature-satellite velocity dispersion relations \citep{zabl98}. A second simulation, Ram/\z{1}{}/Overdensity, adopts f$_{\rm ICM}=1$ to replicate regions of live hydrodynamic models of the cluster medium local to an infalling satellite \citep[e.g.][]{tonn08}, which exhibit a magnitude enhancement in $\rho$ with respect to the spherically-averaged $\rho$ at that radius, and are the preferential regions in which quenched satellites can be located \citep{bahe15}.

Simulation Tidal/\z{1}{}/Harassment denotes the combined results from 16 simulations in which we model the harassment of our infalling disc satellite galaxy by other group satellites (modelled as point massses), in addition to the underlying group tidal field. To appraise the diversity and stochastic nature of harassment, each simulation differs in the prescribed initial location of the infalling satellite. In each case, this location is randomly selected from the locus defined by the group \rvir{}. Although simplistic in its approach, the mean and 1$\sigma$ variation in these simulations' evolution obtained from this Monte-Carlo method provides a useful comparison with our baseline case in determining the systemic effect of harassment.

Simulation Tidal/\z{1}{}/Orientations denotes the combined results from 16 simulations in which the initial inclination of the satellite with respect to its orbital plane is assigned randomly. Again, these results are compared with our baseline cases in which an intermediate infall inclination is adopted, comprising a -45 degree angle between the angular momentum vector of the disc and the group z-axis, a -30 degree azimuthal angle between group x-axis and the projection of the disc angular momentum vector onto the group x-y plane (corresponding to a prograde orbit). 

\figsnapshots

\subsection{Diagnostics}

Simulation data is stored and reduced at 70 Myr increments. The following diagnostics are utilised to compare the evolution of each simulation in terms of transformation and quenching/gas deficiency.

\subsubsection{Morphology}

The stellar/DM/gas mass dynamically associated with the satellite is established with an iterative procedure, wherein gravitationally-bound material discerned in one step is used to build a more refined spherical mass distribution of the galaxy in the following step, until the centre of bound mass is converged upon.

Stellar magnitudes are approximated in this work with synthesis models for the evolution of stellar mass to light ratios \citep{bruz03, port04}. Assuming infall at a lookback time of 8-10 Gyr, an initial stellar age distribution is established from an estimated SF history up to \z{2.5}{}, calculated using extrapolated SFR-stellar relations from \citet{whit12}. We approximate the \hi{}-mass as a factor 1.38$^{\rm -1}$ of each gas particle mass. The attenuation by dust is approximated with the gas column density and an assumed gas-to-dust-extinction ratio of \mex{5}{21}{} cm$^{\rm -2}$A$_{\rm V}$, consistent with bright dwarf/LMC-type galaxies \citep{doba08}.

The surface brightness profile $\mu$($r$) is composed from the corresponding attenuated luminosity, from which a bulge-to-disc mass ratio (\bt{}) is estimated with the excess light lying within 2 effective radii with respect to a fitted exponential disc profile, and normalised with the integrated flux of the disc. The fitted disc profile is estimated with an algorithm that progressively smoothes the radial light profile until the underlying disc (which we find to exist in all cases except where the disc is completely destroyed) can be distinguished from the bulge and/or (positive or negative) truncation components of $\mu$($r$) by lying coincident with a global minimum of d$^{\rm 2}\mu$/dr$^{\rm 2}$.

A B/T$\ge$0.5 is often associated with an early-type morphology \citep{korm04, will09}, although recent studies have highlighted an inherent flaw in adopting simple fits to $\mu$(r) insofar as it incorporates features such as bars/ovals/lenses in the excess light. In the case of early-type systems such as dS0/S0s in which these features are hypothesized to be critical to their formation, this can lead to a significant overestimate of \bt{} \citep{laur09}. 

For this reason, we also compare our satellite's stellar kinematics with those of observed E and/or S0s, using a specific angular momentum $\lambda_{\rm R}$ computed in the manner of \citet{ense07}, who adopt $\lambda_{\rm R}$ to distinguish galaxies as fast/slow rotators:
\begin{equation}
	\lambda_{\rm R}=\frac{\Sigma F_i R_i |V_i|}{\Sigma F_i R_i \sqrt{{V_i}^2 + {\sigma_i}^2}}, 
\end{equation}
where $F_i$, $R_i$, $V_i$ and $\sigma_i$ are the total flux, radius, mean stellar velocity and velocity dispersion respectively, at the $i$th radial bin up to the effective radius (r$_{\rm e}$)

A quantitative demarcation between fast- and slow-rotators can be asserted by the relation $\lambda=0.31\sqrt{\varepsilon}$ found among 260 early-type galaxies in the ATLAS$^{\rm 3D}$ project \citep{emse11}, where the the apparent stellar ellipticity $\varepsilon=1-0.5c/a$, and the axial ratio c/a is computed here from the eigendecomposition of the angular inertia tensor for all bound stars with $r<2r_{\rm e}$. Approximately, $\lambda_{\rm R}\leq0.2$ corresponds to a bright early-type system (E), while S0s and fainter ellipticals can exhibit significant rotational support; this distinction can provide insight as to their formation, featuring less overlap between types than other metrics e.g. V/$\sigma$. 

The formation/survival timescales of bisymmetric structures that are associated with transformation mechanisms are conveyed with two metrics. The bar radius, normalised by the stellar scalelength (r$_{\rm bar}$/r$_{\rm d}$), is estimated via an empirical method \citep[see][and references therein]{yoz14a}; briefly, we detect a bar if a succession of ellipses fit to isophotes of the deprojected, face-on surface brightness distribution (incrementally, from 21 to 28 mag arcsec$^{\rm -2}$) deviate in phase by less than 10 degrees, where the outer most ellipse has an axis ratio no larger than 0.4. 

The disc-wide presence of spiral/tidal arms is quantified with A$_{\rm II}$, the mean amplitude of the second Fourier mode in concentric shells spanning radii from the bar radius up to 2 effective radii; where spiral arms typically register in the range 0.15$<$\asym{II}{}$<$0.6 \citep{rixz95}.

\subsubsection{Star formation}

We express the logarithmic SF rate (SFR) in terms of an offset ($\Delta$SFR) from that expected for a galaxy of its instantaneous stellar mass, for which we adopt the linear relation devised from star-forming galaxies in the SDSS catalog by \citet{kauf03b}. By considering the colour bimodality expressed by the general galaxy population, a passive classification for a galaxy can be determined if $\Delta$SFR$<$-1 (i.e. SFR reduced by a factor 10), with the red sequence lying at approximately -2 \citep[see also][]{bluc14}. A similar result is obtained if we assume that a galaxy can be classified as quenched if its specific SFR (sSFR) $\leq{\rm 10}^{\rm -11}$ yr$^{\rm -1}$, as implied from a bimodality in the sSFRs of central galaxies in the SDSS DR 7 \citep{wetz13}.

Furthermore, the visual signature of SF can manifest in strong \ha{} emission, a tracer of massive star formation which we can approximate with the relation \citep{kenn98}:
\begin{equation}
L(H\alpha)=SFR(M_{\odot}yr^{-1})\times1.26\times10^{41}ergs^{-1}
\end{equation}
where the SFR is estimated from new stellar particles with age less than \mex{3}{7} yr (the approximate age of an $\sim$8 \md{} star).

\subsubsection{Gas mass associated with the satellite}

\figdiagref

We quantify the simulated loss of gas with a \defhi{} parameter \citep{sola96}:
\begin{equation}
	def_{H_{\sc I}}=a + b{\rm log}_{10}(h{\rm D}_{opt}) - {\rm log}_{10}({\rm M}_{H_{\sc I}}),
\end{equation}
where $h$ is the Hubble parameter ($\sim$0.7) and D$_{\rm opt}$ is the linear optical diameter, which we assume equivalent to the galacto-centric radii at which the deprojected azimuthally-averaged surface brightness falls under 25 mag arcsec$^{\rm -2}$. M$_{\rm H_{\sc I}}$ is the total mass of gravitationally bound gas in our simulated galaxy, and $a$ and $b$ are statistically-inferred coefficients which vary according to morphological type. At each time step of our simulations, we classify the type with \bt{} and linearly interpolate between the appropriate coefficients; this, combined with a D$_{\rm 25}$ that evolves according to our adopted stellar synthesis models, yields a \defhi{} parameter comparable with observations.

\section{Results}

\subsection{Dependence of the satellite evolution on group tides since infall at \z{1}} 

Fig. 6 shows the deprojected face/edge-on distribution of stars (in $\mu_{\rm B}$) and gas from our baseline Tidal and Isolated simulations, in a series of snapshots spanning 10 Gyr since their first crossing of the group virial radius. It illustrates how their first pericentre following infall excites the formation of tidal arms and a bar; the associated tidal heating thickens the disc (resulting in boxy isophotes), while the incorporation of an ISM, funnelled to the centre by the bar and tidal torquing, contributes to bulge growth and bar dissipation. Neither tidal stripping nor secular/induced SF can remove the satellites' gas completely within this 10 Gyr timescale.

\subsubsection{Star formation history} 

In the absence of gas accretion. the SF rate among our baseline simulations is shown in Fig. 7 to follow an exponential decline in SFR. This is interrupted only by tidally-triggered SF at each pericentre passage, with an enhancement factor of up to 5 consistent with previous numerical studies \citep{dima08}, and attributed to gas infall from the \hi-rich disc.

The $\Delta$SFR metric implies the mean timescale of 8 Gyr since the accretion of a luminous dwarf galaxy is sufficient for its classification as passive, with a stellar mass growth in the intervening time of only 20 percent. From SFR-mass relations \citep{whit12}, a characteristic SFR of order 1 M$_{\odot}$yr$^{\rm -1}$ betwen \z{0-1}{} for a \ex{9}{} \msol{} galaxy would indicate an analogous isolated galaxy could grow by a factor of 1-5, the majority of which fuelled by a cold gas accretion mode which dominates in this stellar mass regime for all $z$ \citep{deke09}. In conservative analytical models approximating the outflow strength and recycled fraction of gas, the SFR due to fresh accreted gas can be as much as 50 to 90 percent of the total SFR within a gas consumption timescale, for those galaxies hosted by low mass halos in which outflow is relatively efficient \citep{sanc14}. The deficit between the initial SFR of our star forming simulations with respect to $\Delta$SFR$=$0 can therefore be explained by the exclusion of accretion and a gaseous halo from our code, but the subsequent evolution will depend sensitively on the timescale in which this accretion is shut off, which we here assume as instantaneous.

An observable implication of this unreplenished gas reserve, combined with the influence of group tides, is the outside-in truncation of SF (Fig. 8). Previously reproduced by \citet{bekk11} in simulations of massive group spirals and attributed to the heating of the outer disc/gas infall, we extend this result here to dwarf satellites, and indeed find the radial extent of SF typically lies where the stellar/gas Q parameters ($=\sigma\kappa/3.36G\Sigma$, where $\sigma$, $\kappa$, $G$ and $\Sigma$ are the radial velocity disperion, epicyclic frequency, gravitational constant and surface mass density respectively) exceed the canonical range of 1.5-2 required for axisymmetric stability. A similar phenomenon was obtained by \citet{bekk02}, who also attributed the growth in Q and concurrent spiral arm fading with ceased halo gas accretion. 

The ongoing presence of \ha{} emission beyond 8 Gyr ago infall means these satellites (under exclusively tidal influence) would not meet the definition of a quenched system \citep[e.g][]{geha12} nor that of a classical S0; their blue-cores are however comparable with a sample of dEs/dS0s observed in recently accreted subgroups of the Ursa Major cluster \citep{pak14} for which no evidence of significant hydrodynamical influence is found.

\figtoomre

\subsubsection{Gas and halo loss} 

An initial \defhi{} of zero is consistent with our satellite commencing with the mean structural properties/\fgas{} of its mass/type, with a subsequent increase with time corresponding to tidal stripping, outflow and consumption of gas via SF. Uncertainties in \defhi{} among observed galaxies can be large \citep[$\sim$0.25 dex;][]{papp12}, motivating \citet{cort12} to adopt \defhi{}$>$0.5 to denote a deficient galaxy among their sample of 322 nearby galaxies. Adopting this threshold, we find a timescale of 8 Gyr within the group environment sufficient to render an originally gas-rich galaxy \hi-deficient, albeit not removed of gas entirely.

Fig. 9 shows how this reflects a loss of at least 30-60 percent of its original cold disc gas; by comparison, the dark matter halo is truncated such that only 5 percent of its original mass remains bound to the galaxy (Fig. 9); In practice, the dynamical mass evolution within the tidal radius of the satellite (which is dominated by baryons) only amounts to a largely monotonic reduction in time by 50 percent. This result contrasts sharply with those of \citet{wetz13}, who propose a substantial stellar mass growth following group infall as part of a two-stage quenching process wherein the initial stripping of the satellite's halo gas is inefficient.

The gas mass loss exhibited in Tidal simulations includes a secular component which we assume equal to that demonstrated by Isolated; this amounts therefore to about 20 percent loss over 10 Gyr, divided between consumption by SF and outflows. As a sanity check, the corresponding gas depletion timescale of Isolated ($\simeq$50 Gyr) is compared with that expected for its stellar mass, where for the latter we assume a typical HI-to-stellar mass ratio for a \ex{9}{} \msol{} galaxy of 1 to 5 \citep[e.g][]{hayn11} and a median SFR of 0.1 to 0.2 \msol{}/yr \citep{kauf03b}. This yields an acceptable discrepancy by a factor 2-3, broadly attributed to our simple outflow model and the artifical closed-box nature of Isolated.

\figbound

\subsubsection{Transformation mechanisms and efficiency} 

\figprofile

Fig. 10 illustrates the light profile of Tidal/\z{1}{} 8 Gyr after initial infall. In contrast with the single exponential profiles obtained in the collisionless but otherwise similar simulations of \citet{vill12}, our satellites' profiles comprise a bulge component strongly linked with tidally-induced SF \citep{chri04}, and an anti-truncation (up-turn) in the outer disc whose origin is less clear. We find tentative evidence for the break strength $T$ (ratio of inner and outer disc scalelengths) to increase as a function of time resident in the group potential (inset panel). This contrasts with an analysis of the $V$-band $\mu$(r) profiles of 280 classical S0s by \citet{malt15}, a morphological type also preferentially formed in groups \citep{will09} but for which no environmental-dependence of $T$ is found. Instead, \citeauthor{malt15} compare their sample with equivalent spiral galaxies and propose this up-turn is a manifestation of extended bulge light that dominates a fading disc.

The quantitative growth of the central bulges of Tidal/\z{1}{}/Tidal/\z{0}{} and Isolated is shown in Fig. 7. In the absence of merger activity, these are not classical bulges, as further indicated with line-of-sight velocities of the disc which reveal a cylindrical profile in the nucleus. The slow secular growth exhibited by Isolated is consistent with the assertion by \citet{korm04} that pseudo-bulges can comprise up to 10 percent of the total disc mass. By contrast, the bulge growth of the Tidal simulations, whose \bt{} is consistent with the corresponding range, for this stellar mass, of 0$<$\bt{}$<$0.4 collated by \citet{bluc14}, emerges principally as a series of large enhancements coincident with pericentre passages.

The underlying mechanism of this bulge growth is revealed by the formation of a stellar bar, as indicated by r$_{\rm bar}$/r$_{\rm d}$. In both Tidal simulations, the first pericentre passage following infall is sufficient to tidally-trigger this structure to its largest observed radius. A subsequent enhancement in the SF rate and increase in \bt{} is thus consistent with gas infall being faciliated by tidal torques/the stellar bar. However, this simple observable correlation between bar and bulge breaks down with time because the growth of the bulge increasingly dominates the shape of isophotes in the inner disc and thus weakens the visual signature of the bar, whilst the concentration of SF in the bulge means \bt{} can also grow due to the relative fading of the bar and disc. A decline in r$_{\rm bar}$/r$_{\rm d}$ is also to be expected after several pericentre passages due to the buckling of the bar by significant gas infall, similarly reported in simulations of Virgo clusters satellites by \citet{mast05}, and the progressive weakenening of the bar through angular momentum exchange \citep{bou05b}.

\figangmom

An important result however is that the simulated \bt{} is at no point consistent with the range 0.5$\le$\bt{} associated with an early-type morphology \citep{korm04, will09}. This limited morphological evolution, wherein late-type structure is fundamentally maintained, is also reflected in $\lambda_{\rm R}$. In expanding upon Fig. 7, which highlights the sharp loss in angular momentum upon each pericentric passage of the Tidal simulations (with respect to the stability of Isolated), Fig. 11 illustrates the evolution of $\lambda_{\rm R}$ as a function of the apparent stellar ellipticity $\varepsilon$. Their evolution towards lower $\varepsilon$ and $\lambda_{\rm R}$ is consistent with the thickening/heating of the inner disc by tidal shocks upon pericentre passages and the triggered bar. The final kinematics thus ultimately resemble those observed for Sa and S0 morphological types, lying in agreement with our prior determination of its type with \bt{}. Gas accretion from filaments/other galaxies could in principle raise $\lambda_{\rm R}$ if corotating with the stellar disc, although we have assumed no such process occurs within our group enivronment. 

A key feature of these Tidal simulations that precludes their association with the transitory S0 classification is that disc spiral structure beyond the bar proves to be quite robust, being triggered by tides upon all four pericentre passages of the Tidal/\z{1}{} simulation. The peak amplitudes of these arms are short-lived (or order a Gyr) and weaker on successive orbits because i) the stellar disc is increasingly concentrated, and ii) SF that would otherwise be triggered in these outer disc regions is suppressed by the prior infall/stripping of gas and the absence of external gas accretion \citep[e.g.][]{bekk02}. Nonetheless, there is only a $\sim$Gyr window (in Tidal/\z{0}{}) in which our Tidal satellites do not bear the signature of an arm (i.e. where ${\rm A}_{\rm II}<0.15$), strongly suggesting they cannot in general exhibit the axisymmetries inherent to E/S0s.

\subsection{Dependence of the satellite evolution on orbital inclination} 

\figdiagparams

Fig. 12 compares the baseline simulation Tidal/\z{1}{} with the combined results of Tidal/\z{1}/Orientations, wherein we find the former corresponds closely to the mean evolution of the latter in the various metrics that describe the satellite evolution. Variations, in particular those of the triggered starbursts upon first infall, primary stem from how the efficiency with which tidal shocks can trigger bar instabilities/gas infall depend on the disc/orbital plane alignment. On long timescales, however, this dependence becomes ambiguous \citep[e.g. due to stochastic events/resonances of the disc;][]{kaza11}. Moreover, the stripping efficiency of the extended gas disc and DM halo (the latter being prescribed with no initial net angular momentum) are (to first order) invariant with respect to infall inclination.

The clear exception lies in the growth of \bt{} which, although in part a manifestation of tidal heating (due to tidal shocks which prove weakly dependent on inclination), reflects also a sensitivity to the efficiency of bulge growth by the starburst upon first infall \citep{chri04}. Acting upon a near pristine stellar/gas disc, the first group tide interactions that can induce a strong bar/gas infall in a prograde edge-on/inclined scenario (adopted in Tidal/\z{1}{}) are conversely inefficient in a pure face-on and/or retrograde scenario. However, a weak enhancement in \bt{} in these latter cases corresponds to an inefficient manner in which the associated mechanisms described in Section 3.1.3 can destroy/obscure the bar. 

We therefore identify a complex relationship between infall inclination and structural properties of a star-forming group satellite, which manifests in a broad range of \bt{} (since a mean infall lookback time of 8 Gyr) consistent with observations for this mass range \citep{bluc14}. However, the transition to a quenched status over this timescale, reflected in $\Delta$SFR and \defhi{}, remains well described by the reference Tidal models of Section 3.1. 

\subsection{Dependence of the satellite evolution on satellite-satellite interactions} 

Fig. 12 compares the baseline simulation Tidal/\z{1}{} with the combined results of Tidal/\z{1}/Harassment. The principle remark from this comparison lies in the broad scatter revealed by the latter sample; the width of this scatter depends on the interaction softening length, although our adopted constraint is compatible with the typical minimum separation between encounters ($\sim$50 kpc) found by \citet{vill14}, who similarly establish that tidal disruption of a satellite on an infalling orbit is, to first order, governed by interactions with the group halo rather than harassment by other group members.

Satellite encounters at large orbital radii trigger an elevated SF rate upon first infall compared with Tidal/\z{1}{} (and an associated enhancement in \defhi{} also contributed to by tidal stripping), although the SF history largely remains defined by the secular exhaustion of the original gas disc. As noted in Section 3.1.3, the earliest triggers of gas infall/starbursts largely define the bulge mass, as subsequent tidal interactions are unable to trigger as large/strong a bar. These interactions are not sufficient to disrupt the disc directly; instead, three-body interactions or those of unequal mass ratios can either send the dwarf satellite to a higher energy orbit (increased pericentre, period) where the tidal influence of the group halo is weaker, or towards a efficiently disruptive path to the group centre. Two such examples are illustrated in Fig. 12 with a highly divergent evolution in their respective \bt{}. 

\subsection{Dependence of the satellite evolution on interactions with the Intra-Group Medium} %

Fig. 13 compares the evolution in $\Delta$SFR and \fgas{} as exhibited by Isolated and Ram simulations. A key observation here lies in the enhancement in SF wth respect to Isolated following first infall, a phenomenon similarly reported by \citet{bekk14} who attributed this star burst to the compression of the satellites' ISM at the ICM-ISM interface \citep[see also][]{kron08}. In Ram/\z{1}{}, this enhancement can be long-lived because the stripping of the satellites' cold disc gas is inefficient in this scenario, with a final \fgas{} of 60 percent. 

The evolution of simulation Ram/\z{1}/Overdensity indicates that a localised pocket of overdense ICM is sufficient to remove $\sim$50 percent of the satellite's cold gas disc within only its first orbit of the group. If combined with the gas loss exhibited in Tidal/\z{1}{} (i.e. combining simultaneous tidal and ram pressure mechanisms acting on the satellite since first infall), this satellite can lose all its gas and therefore be effectively quenched within 4-5 Gyr of first passing \rvir{}. 

This result is consistent with an apparent peak in the frequency of satellites being quenched by ram pressure stripping at \z{1}{}, where this frequency is not simply a monotonic function of time \citep{bahe15}. However, our result is largely qualitative insofar as our idealisation of the ICM distribtion of a typical group at \z{1}{} carries some uncertainities. The strong dependence of the quenching timescale on the local ICM density of the satellite is therefore the main result of this section.

In an attempt to verify the gas loss simulated numerically in Ram/\z{1}{}, we draw upon the analytical formulation for the satellite stripping radius (r$_{\rm str}$, at which $\rho$V$^{\rm 2}$ exceeds the orthogonal restoring force of the galaxy disc) of \citet{gunn72}:
\begin{equation}
	\rho{V^2} > \Sigma(r)\frac{\delta\phi(r)}{\delta z},
\end{equation}
where $\Sigma(r)$ is the gas surface mass density at galacto-centric radius $r$, and $\delta\phi/\delta z$ is the gradient of the potential field perpendicular to the disc plane. While idealised for a spiral galaxy with co-axial orbital velocity and angular momentum, \citet{roed06} demonstrate that this method can be extended to an arbitrary orientation such as that adopted here. We compute r$_{\rm str}$ per time-step using the (DM, stellar and gas particle) mass distribution and local $\rho_{\rm ICM}$ of our simulated galaxy, and adopt the maximal $\delta\phi/\delta z$ to account for the warp of the disc with respect to our plane of best fit and other difficulties arising from this analytical method as discussed in \citet{roed05}. By applying this r$_{\rm str}$ to the initial structural definition of the gas disc, we compute an analytical \fgas{} from the accumulated gas mass lying beyond r$_{\rm str}$, which we assume instantaneously removed.

Illustrated in Fig. 13, this analytical \fgas{} slightly overestimates the gas loss simulated in Ram/\z{1}{}. This partly reflects the capacity for stripped gas to not actually exceed its escape speed, but more pertinently stems from an assumption in the analytical method for a maximal ram pressure $\simeq\rho$V$^{\rm 2}$, where $V$ is the satellite velocity V$_{\rm sat}$. In practice, the ICM envelope is not stationary, and the galaxy velocity and relative ICM-galaxy velocity can be strongly correlated. A similar result was reported by \citet{tonn08} who establish, for a satellite pericentre velocity V$_{\rm sat}$ of $\sim$500 \kms{} (comparable to Ram/\z{1}{}) in a cluster-scale environment, the ICM ahead of a satellite's trajectory can be travelling at a rate of as much as 0.5V$_{\rm sat}$.

\figram

\section{Discussion and conclusion}

In Section 2, we developed a representative model of a gas-rich late-type dwarf galaxy (stellar mass $\simeq{\rm 10}^{\rm 9}$ \msol), using scaling relations and the mean properties of isolated galaxies in this mass regime, to match their observed mass distributions and SFR as a function of gas surface density. In Section 3, we establish the typical chemodynamical evolution of this satellite following first infall to a (\ex{13}{} \msol) group environment in a suite of simulations. We address the idealistic nature of the group model (where significant mass assembly is not modelled) by characterising the satellites' evolution as bounded by two group models/infall orbits at \z{1}{} and \z{0}{}. 

We first establish, in a pure tidal model of a group (Section 3.1), that the mean time in which a low-$z$ satellite of this mass has been resident with a group ($\sim$8 Gyr) is sufficient to suppress SF to a passive (and \hi-poor) status, if assuming rapid/instantaneous removal of external gas sources (Fig. 6). This galaxy is not quenched, insofar as SF is ongoing (but centrally concentrated); tidal heating plays an active role although the morphological transition to an early type is not complete within a 8-10 Gyr timescale.

In sections 3.2-3, we take a Monte-Carlo approach to assert the sensitivity of this evolution to a variety of orbital inclinations and satellite harassment (Fig. 11), and confirm that the above result is indeed sufficiently characteristic of a satellites' typical evolution. There exists however a broad scatter, in particular for those simulations in which other group members are included, with the implication that the characteristic quenching/transformation can be mitigated for a significant proportion of infalling satellites whose orbits acquire significant energy. 

In section 3.4, we simulate numerically the hydrodynamical ISM-ICM interaction (ram pressure stripping) with results that are consistent with a classic analytical model (Fig. 13), albeit with deviations that can be explained in terms of a correlation of satellite and local ICM velocities. We find in general an enhanced starburst upon first infall, followed by a suppression whose rate depends on the local ICM density; if matching the density to that preferentially found around quenched satellites of previous numerical studies, we find a significant gas loss imposed by ram pressure stripping, with a corresponding quenching of SF within 4-5 Gyr of first infall at \z{1}{}. 

To conclude, our simulations are consistent with observational evidence for the inefficiency with which a \ex{9}{} \msol{} satellite is quenched in a (\ex{13}{} \msol) group environment \citep{whee14}, if most such satellites maintain their relatively slow orbital decay following satellite harassment, and are not resident within a significant local overdensity of the intra-group medium. Our simulations are however idealised, albeit of far higher resolution than previously documented cosmological-scale simulations whose datasets encompass this scenario, motivating the following discussion with proposals for future refinements to our model.

\subsection{A comparison of group tidal mechanisms with previous studies}

Our study is novel with respect to the adopted satellite/host masses, but is otherwise consistent with previous numerical studies that demonstrate how satellites on eccentric orbits of a massive neighbour with weak dynamical friction experience a succession of impulsive tidal shocks (due to the group tides) which heat the stellar disc \citep{gned99}. This mechanism will depend on the orbital pericentre radius/period \citep{mcca08, vill12}, with those satellites that generally avoid the densest (central) regions of their respective environment, such as those modelled in this study, losing their disc (and rotational-support) least efficiently \citep{mast05}.

An integral part of this tidal transformation is the triggering of a bar instability \citep{kaza11}. This is quite resolution-dependent, in light of the stochastic nature of intrinsic disc evolution in numerical models \citep{sell09}, so the robustness of our Isolated simulations to this instability over 10 Gyr provides some confidence in our modelling. We predict therefore the tidal triggering of a bar in a \ex{9}{} \msol{} satellite on its most common infall trajectory \citep{vill12}. The efficiency with which stellar bars are destroyed by the subsequent growth of a central mass concentration/angular momentum exchange from infalling gas is also resolution-dependent, but our results (a bar lifetime of order 1-2 Gyr) are consistent with higher-resolution studies which consider this problem in detail \citep{bou05b}. 

Assuming therefore the slow strangulation of a \ex{9}{} \msol{} group satellite, tidal heating facilitated by our shorter-lived bar will be therefore less efficient than that demonstrated in pure collisionless simulations, such as those of \citet{kaza11} who assumed the early and complete removal of gas in their models of dSph formation from \ex{7}{} \msol{} progenitor discs.

Our approach to modelling a range of harassment scenarios by other satellites (Section 3.3) does not highlight any significant systematic variation to this evolution. A similar result was found by \citet{vill14} for more massive satellites, although their explicit modelling of a live group halo introduced the concept that other satellites can indirectly mitigate satellite transformation by in fact modifying the mutual group potential and the associated satellite orbits. 

We assert therefore that for a given infall trajectory, dwarf satellite transformation is to first order dependent on time resident within the group halo. The concurrent suppression of SF faciliated by group tidal mechanisms is inefficient, but an inverse correlation in the dwarf quenched fraction with mass \citep{whee14} is consistent with the earlier infall times for lower mass satellites \citep{delu12, wetz13}, which would experience a smaller dynamical friction timescale in the lower mass hosts to which they are preferentially accreted at higher $z$ \citep[e.g.][]{jian08}.

This assumes however that we have properly characterised the tidal interaction history of a \ex{9}{} \msol{} satellite i.e. the incidence of major mergers, which we do not model in detail but which can significantly disrupt the stellar system (and form a slow rotating system; Fig. 9) is low. If adopting the definition of a major merger utilised by \citet{rodr15} (e.g. with mass ratio $>0.25$), and integrating their estimated merger rate for this stellar mass which they deduce from the tracking of subhalo merger trees in the Illustris cosmological-scale simulation, we find up to 25 percent of such systems, a minority, will be involved in a major merger since \z{2}{}. This is consistent with most slow rotating systems, whose formation is linked with major mergers, being more massive systems than considered here \citep[$\geq{\rm 10}^{\rm 10.5}$ \msol{};][]{emse11}, and whose mergers rate are higher for all $z$. Moreover, bright dwarves corresponding to our \ex{9}{} \msol{} system are least likely among all dwarves to have been pre-processed and accreted as a satellite to a group-mass host \citep[e.g.][]{wetz15}, a process during which mergers/strong tidal interactions are feasible \citep{gned03}.

The tidal model of the group presented in this study is similar to that of \citet{bekk11}, who demonstrated the transformation of spirals to S0s in response to recent findings that S0s are preferentially formed in ($\sigma<750$ \kms{}) groups \citep[][]{will09, just10}. This supports the role of tidal interactions in their formation, as opposed to the hydrodynamical ISM-ICM interactions which are expected to be dominant in higher-$\sigma$ environments. The long transition timescale conveyed by our satellites is consistent with the continuous growth in S0 numbers since intermediate redshifts \citep{just10}, while the significant reduction in $\lambda_{\rm R}$ and $\varepsilon$ (Fig. 9) is consistent with tidal mechanisms being essential in their formation \citep[as opposed to being simply faded spirals, e.g.][]{vdw10, malt15}.

We stress however that we do not reproduce the classic morphology of S0 in our studies, one discrepancy being the lasting spiral/tidal arm structures in our Tidal simulations (in addition to the ongoing SF which we discuss further in Section 4.2). The inability of our group environment to consistently form S0s is qualitatively in agreement with their being typically more massive than the stellar mass range considered here, as implied by their luminosity functions \citep[e.g.][]{kelv14}. Furthermore, the S0 number fraction peaks in group systems with velocity dispersions of 600-700 \kms{} \citep{just10}, which are nominally far more massive, and quite likely to have accreted from, poor \ex{13}{} \msol{} groups with $\sigma\simeq150$ \kms{} \citep{mcge09}. 

Instead, our Tidal simulations yield systems which are, morphologically and kinematically, characteristic of dwarf S0s. Unlike S0s, these systems are distinguished from the dE classification in which they are often subsumed by a two-component fit to their light profile and, saliently, evidence of a disc including spiral structure \citep{sand84, ague05}. 

Our simulations also convey how tidal influences within a \ex{13}{} \msol{} group will be accompanied with ongoing starbursts that are increasingly concentrated in the disc centre due to gas infall facilitated by tidal torques, bar dynamics and an enhanced $Q$ in the outer disc. Optically, this would replicate the blue cores found among 70 percent of the dE/dS0 galaxies of the Ursa Major cluster by \citet{pak14}, with the dS0 subsample exhibiting the most blue $g-i$ colours for a gven magnitude (in a range $-18<{\rm M}_{\rm r}<-9$). Since this cluster possesses no significant evidence of ongoing ram pressure stripping/\hi-deficiency, they hypothesize that these galaxies represent satellites in transition, whose evolution is primarily driven by tidal interactions. This is supported by a correlation in the frequency of dEs/dS0s with the dynamical state/virialisation of their respective subgroups \citep[see also][]{tull08}, and a systematically more blue colour compared with their counterparts in the Virgo cluster \citep[the latter exhibiting strong evidence of ram pressure stripping;][]{chun07}. Our earlier assertion that the bright dwarves modelled here are least efficiently transformed by group tidal mechanisms is also intriguingly supported by a statistically-significant (at the 3$\sigma$ level) absence of dEs in a range $-18<{\rm M}_{\rm r}<-17$ that closely corresponds to our considered stellar mass. 

\subsection{A comparison of group quenching mechanisms with previous studies}


Our simulations indicate that the tidal mechanism acting within an idealised group (e.g. harassment, stripping) cannot consistently quench a bright dwarf satellite, in which case the environmental processes that dominate the formation of quiescent systems \citep{wetz13}, in particular those in the dwarf mass regime \citep{geha12}, hinge upon the interaction of the satellite with the intra-group medium. A need to  characterise its efficiency is highlighted by increasing evidence that \ex{12-13}{} \msol{} hosts are the building blocks of larger environments, and where quenching mechanisms appear to commence \citep{mcge09, rafi15}.

Our assumption of the instantaneous removal of a \ex{9}{} \msol{} satellites' hot gaseous corona by the group ICM is broadly consistent, albeit conservative, with respect to previous studies which explicitly model this phenomenon and find it almost complete within a Gyr \citep{kawa08, bekk09}. This is supported observationally by the apparent suppression in satellites on their first infall to groups \citep{rasm06}, and even up to several \rvir{} from group centres \citep{rasm12}. The suppression of SF by the shutting off of cold mode gas accretion predicted to contribute $\geq50$ percent to the SFR of bright dwarves \citep{sanc14} is supported further by recent cosmological simulations that highlight how the SF of galaxies is determined by this external gas source prior to infall \citep{rafi15}. Following accretion to halos $<{\rm 10}^{\rm 14}$ \msol{}, \citeauthor{rafi15} demonstrate how the \hi{}-richness of galaxies traces the sSFR, implying the starvation of the satellite on a timescale of several Gyr.

Various studies espouse a similar timescale during which the SFR declines in an exponential manner since infall, with $e$-folding timescales inferred to be of order 1-3 Gyr \citep{wang07, mcge09, delu12} but which in practice can vary widely depending on respective satellite and host masses \citep{bekk14}. This lies broadly in accordance with the predominantly secular consumption of extant disc gas in our Tidal simulations irrespective of triggered starbursts (Fig. 6), which proceeds according to a SF model that matches a dependence on gas surface density for analogous metal-poor dwarf galaxies \citep{bigi08}. The low SF efficiency with respect to e.g. the Kennicutt-Schmidt relation (Fig. 4) is thus demonstrated to match the slow transition to an inactive status.. 

An alternative two-stage mode of suppresion was recently introduced by \citet{wetz13}, who compare the SFR of satellites with a statistical parametrization of central galaxy SFRs in the SDSS DR 7 (with a completeness limit lying slightly above our stellar mass at \ex{9.7}{} \msol{}). Their analysis indicates that the suppresion of SF in dwarf satellites can be delayed by more than 4 Gyr after infall (during which the build up of new stellar mass is substantial, of order 50 percent), followed by a rapid decline in SFR with timescale $<1$ Gyr. Their conclusion that these satellites are accreted with enough cold gas to sustain the SFR prior to infall is not supported by our simulations, although we have adopted an initial gas fraction (\fgas{}) implied by observations for this stellar mass \citep[$\sim1$;][]{popp14}. This may underestimate the total associated gas, which has recently demonstrated for infalling satellites of the Illustris simulation to be in the range f$_{\rm gas}=2-8$ \citep{sale15}.


We do note however a quantitative agreement in this two-stage satellite quenching with our ram pressure simulations (Section 3.4). Fig. 13 conveys a high sustained SFR (for several Gyr) in the Ram/\z{1}/Overdensity simulation, before the available gas is exhausted and/or the local ram pressure at pericentre passage is sufficient to effectively quench the satellite within 5 Gyr of first virial crossing. This triggering of SF at the ICM-ISM interface, with simultaneous stripping in the outer disc, contributes to the concentration of SF at its centre where the restoring force is largest \citep{bekk14}. A similar enhancement in SFR and a truncation in the \ha{} scalelength, compared to field counterparts, was noted in analyses of Virgo cluster spirals \citep{koop04, koop06}. This phenomenon is not limited to dense environments, as revealed by the observation of both a starburst on the leading edge and rapid gas loss via ram pressure stripping in the trailing tail of a late-type satellite of a poor group by \citep{rasm06}. On the other hand, an analysis by \citet{rasm12} of SF rates among the constituents of 23 local galaxy groups finds only 1 to 10 percent of group galaxies exhibiting evidence of starbursts; instead, the net effect on SF, with respect to their field sample, is a suppression by $\sim$40 percent at group-centric distances up to 2\rvir{}.

A sustained triggering of SF is limited to a model in which we assume a locally overdense ICM, which recent large-scale hydrodynamical simulations predict infalling satellites to preferentially lie within, with the corresponding quenching timescale in group hosts less than the first orbital period \citep{bahe15, cen14}. However, their results, implying the complete quenching of the \ex{9-9.5}{} population of satellites of \ex{13-13.5}{} \msol{} groups within 4 Gyr contrast sharply with those of \citet{whee14}, who adopt mock observations from the Millenium-II simulation and establish the long quenching timescales ($\geq6-7$ Gyr) that motivated this paper. At present, we emphasise that these cosmological simulations utilise baryonic particle masses of order \ex{7}{}h$^{\rm -1}$ \msol{} (corresponding to a resolution of only $\sim$100 particles for a stellar mass of \ex{9}{} \msol{}). Nonetheless, we identify some promising agreement with our better resolved but idealised group infall model that encourages further refinement.


\section*{Acknowledgements}

We thank the referee for their insightful comments which led to the improvement of our manuscript. CY is supported by the Australian Postgraduate Award Scholarship. The numerical work reported here was conducted on three GPU clusters (Pleiades, Fornax and gSTAR) kindly made available by the International Centre for Radio Astronomy reseacrh (ICRAR) at the University of Western Australia, iVEC, and the Centre for Astrophysics and Supercomputing at Swinburne University, respectively. This research was supported by resources awarded under the Astronomy Australia Ltd's ASTAC scheme at Swinburne with support from the Australian government. gSTAR is funded by Swinburne and the Australian Government's Education Investment Fund.

\bibliographystyle{mn2e}
\bibliography{bib}

\bsp
\label{lastpage}
\end{document}